\documentclass[aps,amsfonts,amsmath,amssymb,nofootinbib,twocolumn, superscriptaddress]{revtex4-2} %linenumbers

\usepackage{graphicx}
\usepackage{bm}
\usepackage{hyperref}
\hypersetup{
    colorlinks,
    linkcolor={red!50!black},
    citecolor={blue!50!black},
    urlcolor={blue!80!black}
}
\usepackage{float}
\usepackage[dvipsnames]{xcolor}
\usepackage{float}
\usepackage{ulem}
\usepackage{titlesec}
\usepackage{pdfpages} 
\usepackage{etoolbox} 

\makeatletter
\patchcmd{\@outputpage@head}{\@ifx{\LS@rot\@undefined}{}{\LS@rot}}{}{}{}
\makeatother

\titleformat{\section}{\large\bfseries}{\thesection}{1em}{}

\graphicspath{{figures/}}
\let\comma,

\begin{document}

\title{Decoupling momentum and energy relaxation rates in cuprate strange metals via giant THz nonlinearities}

\author{Dipanjan Chaudhuri$^{\dagger}$}
\author{David Barbalas$^{\dagger}$}
\affiliation{William H. Miller III Department of Physics and Astronomy\comma{} Johns Hopkins University\comma{} Baltimore\comma{} Maryland 21218\comma{} USA}

\author{Fahad Mahmood}
\affiliation{William H. Miller III Department of Physics and Astronomy\comma{} Johns Hopkins University\comma{} Baltimore\comma{} Maryland 21218\comma{} USA}
\affiliation{Department of Physics\comma{} University of Illinois Urbana-Champaign\comma{} Urbana\comma{} Illinois 61801\comma{} USA}
\affiliation{Materials Research Laboratory\comma{} University of Illinois Urbana-Champaign\comma{} Urbana\comma{} Illinois 61801\comma{} USA}

\author{Jiahao Liang}
\author{Ralph Romero III}
\author{Ana\"elle Legros}
\affiliation{William H. Miller III Department of Physics and Astronomy\comma{} Johns Hopkins University\comma{} Baltimore\comma{} Maryland 21218\comma{} USA}

\author{Xi He}
\affiliation{Brookhaven National Laboratory\comma{} Upton\comma{} New York 11973\comma{} USA}

\author{H\'el\`ene Raffy}
\affiliation{Laboratoire de Physique des Solides (CNRS UMR 8502)\comma{} Universit\'e Paris-Saclay\comma{} 91405 Orsay Cedex\comma{}  France}

\author{Ivan Bo\v{z}ovi\'c}
\affiliation{Brookhaven National Laboratory\comma{} Upton\comma{} New York 11973\comma{} USA}
\affiliation{Shanghai Advanced Research in Physical Sciences (SHARPS)\comma{} Pudong\comma{} Shanghai 201203\comma{} China}

\author{N.\,P. Armitage}
\email{npa@jhu.edu}
\affiliation{William H. Miller III Department of Physics and Astronomy\comma{} Johns Hopkins University\comma{} Baltimore\comma{} Maryland 21218\comma{} USA}
\affiliation{Canadian Institute for Advanced Research\comma{} Toronto\comma{} Ontario M5G 1Z8\comma{} Canada}
 
\maketitle

\def\thefootnote{$\dagger$}\footnotetext{These authors contributed equally to this work.}

\textbf{Understanding the $T$-linear normal-state resistivity of cuprates remains a central physics challenge. The associated momentum relaxation rate, $\Gamma_M$, saturates near the conjectured ``Planckian" bound $\Gamma_M\sim kT/\hbar$, but the mechanism underlying the anomalous scattering remains unresolved. Here we employ nonlinear terahertz spectroscopy to systematically study La$_{2-x}$Sr$_x$CuO$_4$ across a broad temperature and doping range. We measure the normal-state third-order susceptibility, $|\chi^{(3)}|\approx 6\times10^{-9}$ m$^2$/V$^2$, among the largest in the THz regime, enabling direct access to the rarely measured electronic energy relaxation rate, $\Gamma_E$. Strikingly, $\Gamma_E$ is 10-40 times smaller than $\Gamma_M$, revealing that the scatterings responsible for momentum loss and $T$-linear resistivity do not remove appreciable energy from the electrons. While $\Gamma_M (T)$ is consistent with quasi-elastic scattering from bosonic modes above their characteristic energy scale, this is incompatible with the increasing temperature dependence of $\Gamma_E(T)$.  Our results exclude phonons as the source of $T$-linear resistivity and impose strong constraints on possible mechanisms.}

\medskip

\section*{Introduction}

The normal state of the cuprate superconductors continues to be among the most enigmatic problems in physics even after almost four decades of intense investigation \cite{anderson2013twenty, phillips2022stranger}. Near optimal doping, the system exhibits a $T$-linear resistivity, maintiaing a constant slope from low temperatures to well above room temperature. Assuming a $T$-independent carrier density and mass, this suggests a scattering rate $\Gamma \sim kT$. While $T$-linear resistivity due to quasi-elastic electron-phonon scattering can occur in normal metals over an extended temperature range above the Debye temperature \cite{hwang_linear2019}, it is expected to saturate when the mean free path becomes as short as a lattice constant.  The existence of $T$-linear resistivity over a large range is believed to signify a non-Fermi liquid regime without well-defined quasiparticles.  Despite extensive experimental studies \cite{gurvitch_resistivity_1987, hussey_generic_2013, legros_universal_2019, grissonnanche_linear-temperature_2021,cooper_anomalous_2009,daou2009linear}, the microscopic origin of this \textit{strange metal} state remains unclear.

Developing new experimental probes that are selective to particular kinds of scattering may help in parsing the strange metal state of the cuprates.   It has recently been shown that nonlinear THz 2D coherent spectroscopy (2DCS) on interacting metals can be particularly informative, if one decomposes the polarization dependence \cite{barbalas2025energy} of the ``pump-probe" contributions to the 2DCS response.  The decay rate of the isotropic contribution can be connected to the energy relaxation rate \cite{allen1987theory, glorioso2022joule} i.e. the rate at which energy leaves an excited electronic system.  The study of energy relaxation dynamics in correlated metals may assist in determining the contributions to the transport scattering rate, the relative importance of elastic and inelastic scattering, or the effects of isotropic and anisotropic scattering around the Fermi surface.  Such THz experiments avoid the issues with the high photon energies ($\hbar \omega \sim $ eV) used in conventional optical pump-probe experiments, which can be affected by energy dependent scattering~ \cite{kabanov_kinetics_2005}, excite optical phonons, induce phonon bottlenecks, and otherwise drive far-from equilibrium that can be challenging to model [see \textit{Supplementary Information}].  %Recent developments in intense THz sources have enabled nonlinear spectroscopy with low-energy THz photons that only excite quasiparticles near the Fermi surface \cite{barbalas2025energy}. 
More broadly, the THz 2DCS technique is well-suited to study coherent excitations \cite{lu_coherent_2017, johnson_distinguishing_2019}, superconducting properties \cite{luo2023quantum, liu2024probing}, order parameter dynamics \cite{matsunaga_nonequilibrium_2012, matsunaga_light-induced_2014, pal_origin_2021}, and population dynamics \cite{hebling_observation_2010, mahmood_observation_2021}, among other phenomena \cite{liu2025multidimensional, kuehn2011two, huang2025unlocking}. Applications of THz 2DCS in correlated metals have provided new insights, as it can offer a complimentary probe to transport experiments that measure the momentum relaxation rate \cite{barbalas2025energy,bhandia2024anomalous}.

La$_{2-x}$Sr$_x$CuO$_4$ (LSCO) is a prototypical hole-doped cuprate with a single CuO$_2$ plane per unit cell.
%one LSCO layer that is about 0.62 nm thick.  
Superconductivity onsets at the Sr level of $x\approx0.05$ and exhibits the maximum $T_c\approx$ 40~K at an optimal doping of $x \approx 0.16$. Further increase in doping results in a decrease of $T_c$ until superconductivity disappears near $x\approx0.3$ and the material enters what appears to be a more conventional metallic phase. The normal state of the underdoped cuprates ($x<0.16$) hosts a number of charge- and spin-ordered states, whereas the overdoped ($x>0.16$) side has been regarded to be more conventionally metallic \cite{keimer1992magnetic}.

In this work, we report the observation of a large nonlinear THz response in LSCO films. The nonlinearity is largest at low temperatures near optimal doping levels in the superconducting state, but persists to temperatures well above $T_c$. This is in sharp contrast to that of metals and conventional BCS superconductors where no measurable THz nonlinearity is observed in the normal state.
%NbN where the nonlinearity promptly disappears at $T_c$, or that of the normal metal Au where the effect is orders of magnitude smaller at this frequency range with the same field strengths. 
For an overdoped sample well above $T_c$, the estimated third-order susceptibility, $|\chi^{(3)}|\approx (6\pm2)\times10^{-9}$m$^2/$V$^2$, which is one of the largest nonlinearities ever measured in the THz range. The anomalous nonlinearity may stem from strong interaction effects in the strange metal, or from precursor superconductivity extending up to 40~K above $T_c$.  THz 2DCS reveals that pump-probe signal persists across $5-100$ K, which, in the normal state is governed by the energy relaxation ($\Gamma_{E}$) out of the electronic system.  Over this range, $\Gamma_E$ increases with temperature and but remains $10-40$ times smaller than the momentum relaxation rate, $\Gamma_{M}$.  This suggests that the scattering processes that degrade momentum largely do not remove energy from the electronic system.  Although this could be consistent with electron-bosonic scattering at temperatures above the characteristic energy scale of the bosonic bath (e.g. phonons above their Bloch-Gr{\"u}neisen temperature), leading to quasi-elastic scattering, it is inconsistent with the observed increase of $\Gamma_{E}$ over the same temperature range \cite{allen1987theory}.    Quasi-elastic scattering would exhibit a $\Gamma_{E}$ that is a decreasing function of temperature.  Among other aspects, this rules out phonons as a source of $T$-linear resistivity in the cuprates.   What other possibilities it rules out remains an open question.

\section*{Results and Discussion}
%\subsection*{Experimental Details}

We measured the nonlinear THz response for 22 MBE-grown LSCO thin films across a wide range of doping levels, ranging from strongly underdoped ($T_c=7$ K) to non-superconducting overdoped ($T_c \approx 0$ K) samples [see \ref{Methods}{Methods}].% The films were grown by molecular beam epitaxy (MBE) on LaSrAlO$_4$ (LSAO) substrates.  Growth was monitored \textit{in-situ} in real-time using RHEED.  The films were additionally characterized using x-ray diffraction.  Two-coil mutual inductance was used to measure $T_c$.  
Due to oxygen non-stoichiometry the Sr content does not directly correspond to the hole content of the CuO$_2$ planes. We have used the formula $T_c/T_{c,max} \approx 1-82.6(p-0.16)^2$, to extract the nominal hole concentration, $p$, but this expression is at best empirical~\cite{presland1991general}. The measurements were carried out in a now standard setup for THz 2DCS [see \ref{Methods}{Methods}, \cite{mahmood_observation_2021, barbalas2025energy}]. %Intense THz pulses were generated by exciting a pair of LiNbO$_3$ crystals with amplified laser pulses at 800 nm with $\approx$3 mJ energies in the tilted-pulse-front scheme~\cite{hirori2011} and detected with standard electro-optic methods using ZnTe (for the single-pulse transmission experiments) or GaP (for all other two-pulse measurements) detection crystals. 
We achieved maximum field strengths of 110 kV/cm, which were typically strongly attenuated with wire-grid polarizers to study the response in a perturbative regime. Most measurements employed a two-pulse scheme ($A$ and $B$), in which both the inter-pulse delay $\tau$ and the lab detection time $t$ can be varied.  Nonlinearities are isolated as $E_{NL}(\tau,t) = E_{AB}(\tau,t) - E_A(t) - E_B(\tau,t)$.  %Further details of the THz setup are reported in Refs. ~\cite{mahmood_observation_2021, barbalas2025energy}.   
Experiments on bare LSAO substrates showed no measurable nonlinearities up to the highest $E$-fields.  These results rule out contributions from the substrate as well as artifacts arising from nonlinearities in the detection crystals \cite{liu2024excitation,selz2025terahertz}.  %Pumping and probing with co- and cross-polarized geometries was performed, but 
No measurable polarization anisotropy was observed either [\textit{Supplementary Information}].

\begin{figure}
    \includegraphics[width = 1.05\columnwidth]{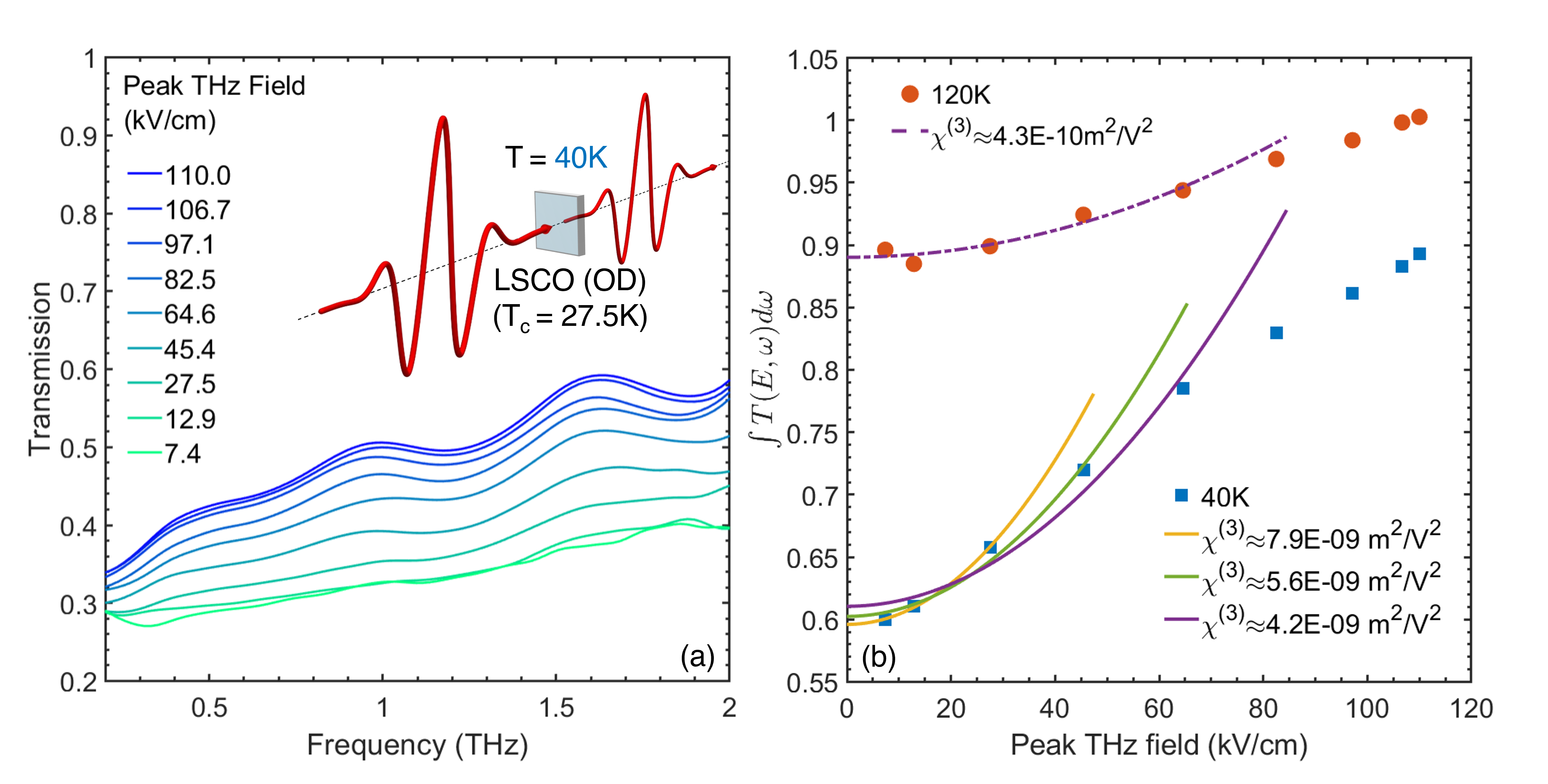}
    \caption{\label{Transmission} (a) Transmission of overdoped LSCO ($T_c$ = 27.5 K, $x \simeq 0.23$) as a function of frequency for different incident $E$-fields at 40~K normalized by the substrate transmission. (inset) Schematic of the experiment. (b) Solid (dashed) lines represent quadratic fits over different ranges of the integrated transmission at 40 K (120 K), from which the corresponding $\chi^{(3)}$ coefficients are extracted.}
\end{figure}

\begin{figure*}
    \includegraphics[width = \textwidth]{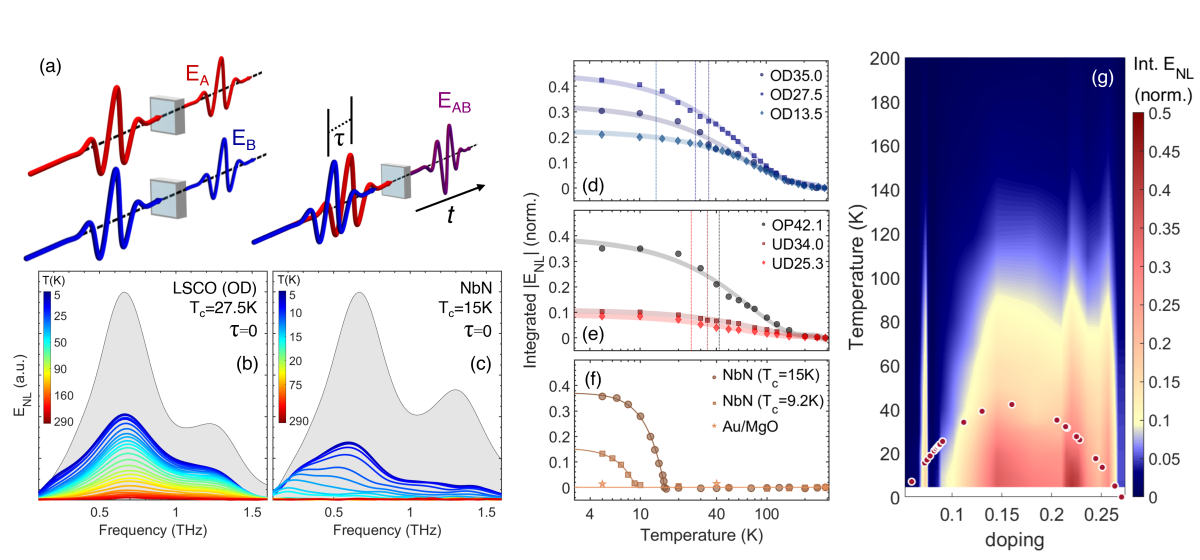}
    \caption{\label{Comparison} (a) Schematic of the two-pulse experiment. The transmission from two THz pulses, $E_A$ and $E_B$ are first measured independently, followed by the transmission of the two applied simultaneously, $E_{AB}$. The difference in the transmitted THz pulses yields the nonlinear response $E_{NL}$. (b) The spectrum of the nonlinear response from overdoped LSCO ($T_c$=27.5~K) (c) and a $s-$wave superconductor NbN ($T_c$=15~K). The grey shaded area in (b,c) shows the spectrum of the total incident THz pulses.
    Amplitude of frequency integrated nonlinear signal from (d) overdoped, (e) optimal, and underdoped LSCO thin films. Solid vertical lines denote the $T_c$ for each sample. (f)  The same for clean NbN, and weakly disordered NbN films, and a gold film on MgO substrate. The response is normalized by $|E_A+ E_B|$ to account for the changes in linear response.  The nonlinear response in NbN can be fit to a phenomenological BCS gap equation. (g) Color map of the relative nonlinear response for various dopings of LSCO. Red dots indicate the $T_c$ of the samples measured.  Data was taken with co-polarized pulses.  }
\end{figure*}

\subsection*{Experimental results on normal state nonlinearities}

The nonlinearity in the THz response can be probed by transmission measurements performed at different $E$-field strengths (Fig.~\ref{Transmission}). Within linear response theory, the normalized transmission is a material property, \textit{independent} of the probing field strength.  In contrast, we observe a large enhancement in the transmission (normalized to a substrate) with increasing $E$-field. In Fig. \ref{Transmission}(a) we show this effect in an overdoped sample ($x = 0.23$) in the normal state at 40~K ($T_c$ = 27.5 K) where transmission between 0.2-2 THz increases with increasing field strengths. The induced transparency is broadband, and hence not related to any specific resonant absorption process.  While such broadband THz field-induced transparencies have been observed in graphene \cite{hwang2013nonlinear, mics2015thermodynamic}, they were attributed to the linear, Dirac-like dispersion, which is not relevant to LSCO. Here the effect persists up to high temperatures as shown in Fig. \ref{Transmission}(b) where we plot the integrated transmission amplitude $\int T(E,\omega) d\omega$ at 40K and 120K, although the nonlinearity is smaller at higher temperatures.  With increasing field strength, $T(E)$ increases and gradually saturates at high fields, mimicking the saturable absorber response in semiconductors ~\cite{haiml2004optical, hoffmann2010semiconductor}.  It is important to emphasize that this large nonlinearity is a property of the normal state.  Below $\sim$ 45 kV/cm, the field dependence of  $ T(E) $  has the quadratic dependence expected for a $\chi^{(3)}$ nonlinearity.

To explicitly quantify the temperature dependence of the nonlinear response, we employ a two-pulse measurement scheme described previously.% Two intense THz pulses, $E_A$, and $E_B$, are collimated and co-polarized using a set of wire-grid polarizers.  
We use co-polarized, overlapping ($\tau = 0$) intense THz pulses, and directly measure the nonlinear response $E_{NL}(t)$ in the time-domain. The maximum field strength we used here in the individual pulses $E_{A,B}$ was $\approx$ 15 kV/cm.

\begin{figure*}
    \includegraphics[width = \textwidth]{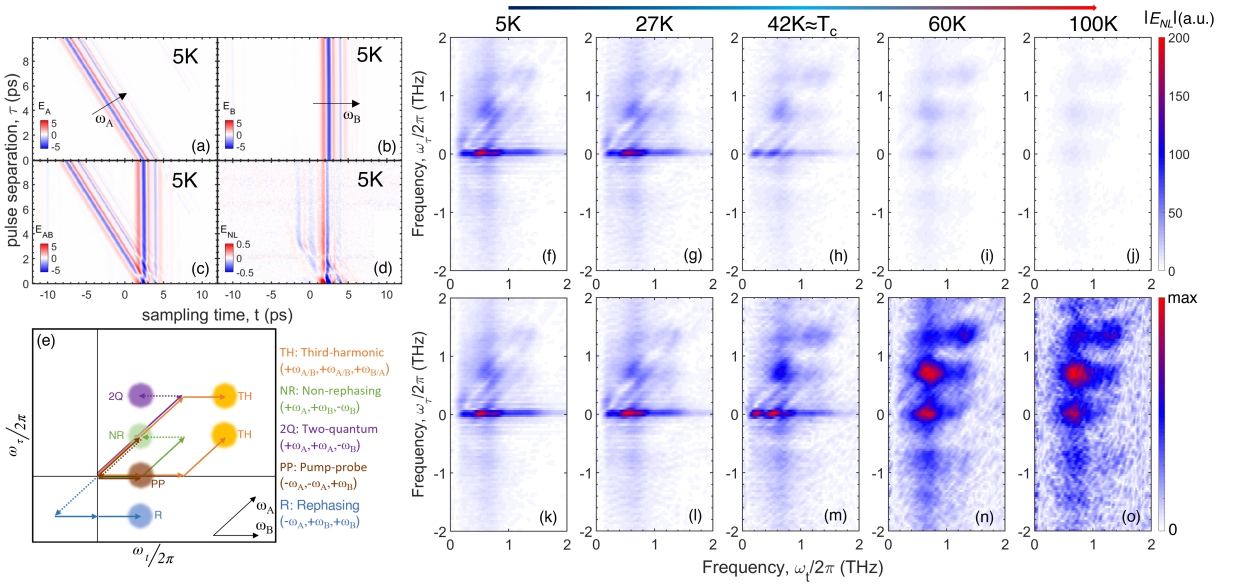}
    \caption{\label{2D} Time-traces of (a) $E_A$, (b) $E_A$, (c) $E_{AB}$, and (d) $E_{NL}$ as a function of pulse separation, $\tau$, for an optimally doped LSCO film ($T_c$=42.1~K) measured at 5~K. (e) Schematic of the frequency-vector scheme for the 2DCS showing the $\chi^{(3)}$ processes for the given pulse sequence.  (f-j) Temperature evolution of the THz 2DCS are shown in absolute color scale. (k-o) Saturated colormap for the same data as in the top panel highlighting the relative strength of various  $\chi^{(3)}$ processes. The pump-probe response dominates the spectrum below $T_c$.  Data was taken with co-polarized pulses and 15 kV/cm.}
\end{figure*}
 
In Fig.\ref{Comparison}(b, c), we show the THz nonlinear response measured in the frequency domain for films of slightly overdoped LSCO and a conventional $s$-wave superconductor NbN.  The LSCO film is the same $T_c = 27.5$K sample discussed above.  It shows a large nonlinear response at low temperatures that smoothly decreases as temperature is increased. This behavior contrasts sharply with that of NbN films, where we observe the anticipated strong nonlinear response below $T_c$, which vanishes rapidly upon entering the normal state. 

To further investigate the temperature dependence, we integrate the amplitude of the nonlinear signal ($\tau = 0$) between 0.1-1.6 THz, normalized to the incident $E$-field $E_A+E_B$. Fig. \ref{Comparison}(d, e) shows the integrated nonlinear response for selected overdoped and optimal/underdoped LSCO films respectively which evolves smoothly across $T_c$ and disappears at temperatures $T \gg ST_c$. This behavior is qualitatively distinct from that of moderately and weakly disordered NbN films as shown in Fig.~\ref{Comparison}(f) where the large integrated nonlinearity, normalized using the same protocol, is only observed in the superconducting state and can be fit to an expression proportional to the BCS order parameter $\Delta(T) = \Delta_0\tanh\left({1.74\sqrt{T_c/T-1}}\right)$. The integrated nonlinear response for a thin gold film deposited on MgO remains below the detection limit at all temperatures.  As mentioned above, the bare LSAO substrate also shows no measurable nonlinearity.

In Fig. \ref{Comparison}(g) we plot the compiled integrated nonlinear response for 22 LSCO films that span the superconducting dome. We estimate the doping based on the empirical relation described above, but it is important to make a distinction between nominal doping, which in this case refers to $x$, the Sr concentration in LSCO (which is well controlled) and the number of electrons per unit cell in the CuO$_2$ plane, $p$.  The latter can depend on several factors including the oxygen stoichiometry which is influenced by parameters such as the ozone partial pressures and cool-down procedures.  One can see that the nonlinearity roughly tracks the domain of superconductivity itself.  The doping level where the nonlinearity persists to the highest $T$ is near the doping level of the highest $T_c$, but skewed more toward the overdoped side.  The nonlinearity decreases and largely vanishes for underdoped and strongly overdoped samples.   Note that as seen in Fig. \ref{Comparison}(g), one very underdoped sample with $T_c$=16.9K showed an enhanced nonlinearity, which is out of the trend.   As this sample also showed large superfluid density, we believe this sample's signal is just spurious and reflects just some large sample-to-sample variation, which is a persistent issue in underdoped cuprates.

To isolate which nonlinear processes are contributing to the large overall nonlinear response, we show in Fig.~\ref{2D} the 2DCS data from the optimally doped sample ($T_c=42.1$K). As discussed above, the data for $E_{NL}$ are taken as a function of $\tau$ and $t$ and then Fourier transformed to get the 2D spectrum. Figs.~\ref{2D}(a-d) show the raw time traces for $E_{A}$, $E_{B}$, $E_{AB}$, and E$_{NL}$ at 5~K for $\tau>0$. As the crystal structure of LSCO preserves inversion symmetry, bulk even-order nonlinear responses are forbidden and the dominant contribution at the lowest order in $E$-field is a first-harmonic $\chi^{(3)}$ process. Such responses can arise from pump-probe ($PP$), rephasing  ($R$), and non-rephasing ($NR$) pathways \cite{kuehn2011two, lu_coherent_2017, mahmood_observation_2021}.  THz 2DCS separates these contributions by mapping them to distinct regions of the 2D spectrum (Fig. \ref{2D}(e)). 
Measurements at various temperatures [Fig. ~\ref{2D}(f–o)] exhibit no sharp modes, but a broad response with contributions from multiple regions of the 2D map.

Below $T_c$, the most prominent peak is along the horizontal axis of the 2D map. Based on the measurement configuration and the inclusion of only time-domain data for $\tau>0$ in the Fourier transform, this is where we expect to observe the $PP$ signal.  In systems with discrete energy levels, this signal measures a response governed by population decay \cite{mahmood_observation_2021}.   The dominant contribution of the $PP$ signal below $T_c$ may be a result of being in the non-perturbative regime already at 15 kV/cm in the superconducting state.  Kuehn et al. had shown in a semiconductor that increasing excitation field strength beyond the perturbative $\chi^{(3)}$ regime results in a dominant pump-probe response, just as observed at present in the superconducting state~\cite{kuehn2011two}.

However, the normal state data appears to be safely in the perturbative regime.  Above $T_c$, broader $NR$ and $R$ contributions seem almost comparable to the pump-probe response.  Whether these are intrinsic and perhaps a consequence of strong interactions, or arise from residual $B$ before $A$ contributions (that put $PP$ signal into the position where $NR$ would appear for $AB$ ordering~\cite{mahmood_observation_2021}) will be a subject of future investigation.  We expect the $PP$ response to be the dominant low-frequency nonlinear response in metals where only $\omega \rightarrow 0$ excitations are present. For example, ``echo" phenomena (e.g. $R$) are not possible in a such a system as the effective Larmor frequency is zero.  At any rate, these $R$ and $NR$ spectral contributions are broad and therefore do not contribute to the long-time $\tau$ signal that we analyze below.

\subsection*{Discussion on normal state nonlinearities}

The magnitude of the $\chi^{(3)}$ nonlinearity can be determined by fits to the quadratic field dependence of the transmission.   The expression is 
\begin{eqnarray}
   | T_{eff} | = | T_{lin} |\Big[1 + \Big | \frac{ \omega\epsilon_0 \chi^{(3)} E_{inside}^2}{\sigma^{(1)}}  \Big | + \; \;...\Big]
\end{eqnarray}
where $ T_{lin}$ is the linear transmission at small fields and $E_{inside}$ is the driving field corrected by screening effects (see the \textit{Supplementary Information} for details).

Based on fits to the THz field-induced transparency shown in Fig.~\ref{Transmission}(b), the third-order susceptibility can be estimated to be $|\chi^{(3)}|\approx (6 \pm 2 )\times10^{-9}$m$^2/$V$^2$.  This is one of the largest THz range nonlinearities ever measured and approximately the same order as that of graphene, already known to be one of the most nonlinear materials in the THz range~\cite{hafez2018extremely}.   The error bars were set by varying the range of the quadratic fit.   Alternative methods for estimating the nonlinear coefficient by including fitting higher-order terms and a larger fitting range are detailed in the \textit{Supplementary Information} and gave values consistent with those reported here.

\begin{figure}
\begin{center}
    \includegraphics[width = 0.95\columnwidth]{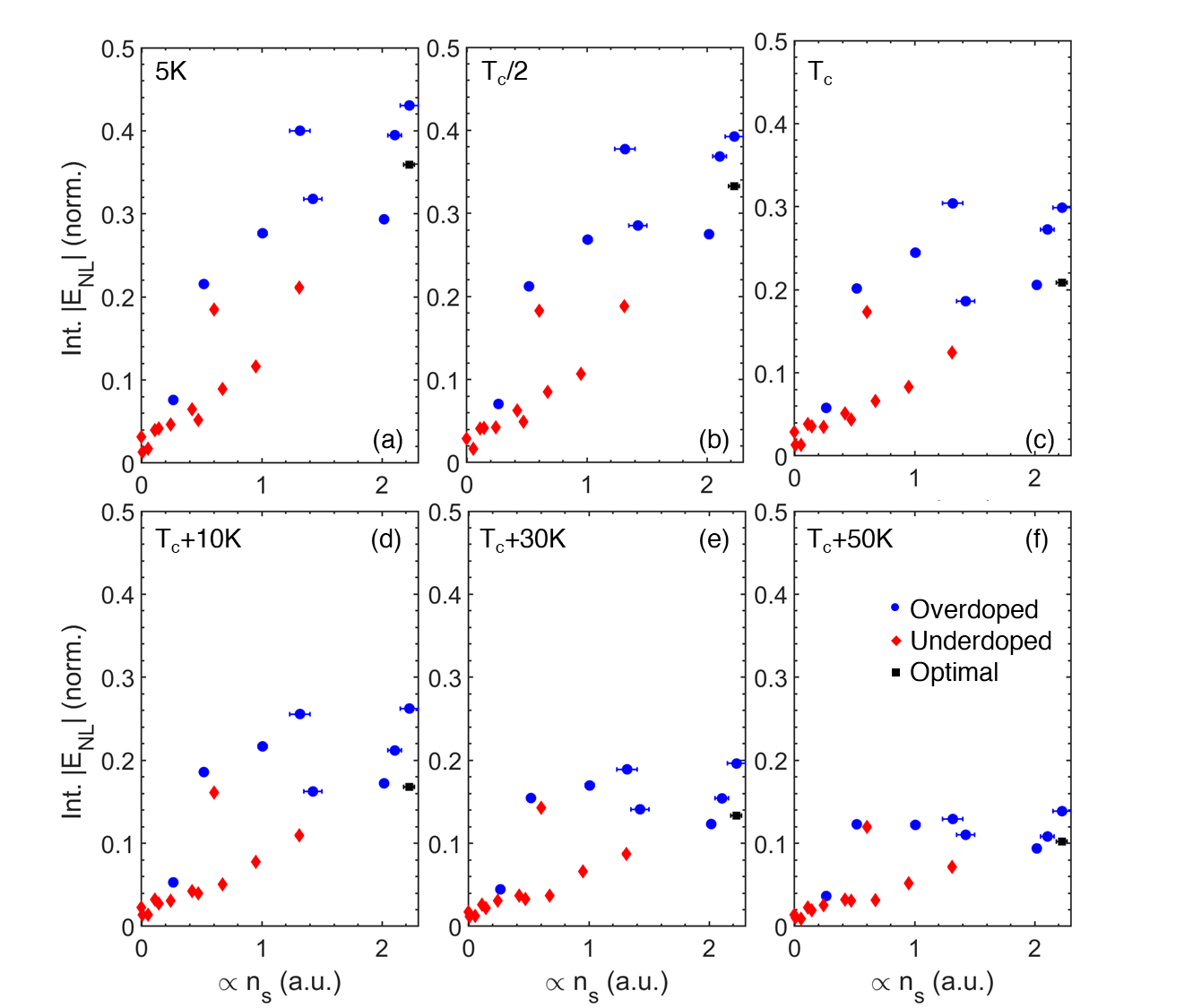}
    \caption{\label{ns_NL} The integrated nonlinearity ($\tau = 0$) for various samples plotted as a function of the low temperature (T$\ll T_c$) superfluid density of the samples measured in linear response. Each panel represents the nonlinearity at various temperatures relative to $T_c$. A positive correlation is observed between THz nonlinearity and the superfluid density, even for T$>$ $T_c$.}
\end{center}
\end{figure}

Several mechanisms could be responsible for such large THz nonlinearities. A metal with parabolic bands and only energy-independent elastic scattering is expected to exhibit any nonlinearities. THz nonlinearities arise if the bands are non-parabolic (as in graphene, in the extreme case) or scattering is energy-dependent \cite{rustagi_dispersion_1984, markelz_relaxation_1998}. Heuristically, transport nonlinearities emerge because electrons excited to large, non-equilibrium momenta experience different effective mass and/or scattering than at low momenta. Thus, it is reasonable to expect that strongly interacting metals, such as cuprates, could exhibit pronounced low-frequency transport nonlinearities. 

Recently, Kryhin et al. [Ref. \cite{kryhin2024strong}] pointed out that strong THz nonlinearities are likely to be a generic feature of the strange metal due to strong scattering. Because scattering in the strange metal scales as $kT$ instead of as $kT^2/E_F$ as in a Fermi liquid, one expects enhancements of nonlinearities over that of an analogous Fermi liquid by a factor of order $E_F/kT$. Note that this is qualitatively consistent with our finding as both Au and NbN showed no normal state nonlinearities within the experimental uncertainty, but $E_F/kT$ should be a large factor greater than 10$^3$ in the present case. In this respect, it is notable that large nonlinearities on the frequency scales relevant for transport are a {\it generic} outcome of the strange metal state in the cuprates. 

The nonlinearity could also arise from nonparabolicity, which is particularly enhanced near optimal doping in LSCO. %It has been established that there exists 
A van-Hove singularity occurs near $p\sim$0.19 where the Fermi surface changes from hole-like to electron-like -- very close to the doping level where the maximum nonlinearity is found (Fig. \ref{Comparison}(g)). Near this ``Lifshitz transition", nonparabolicity in the bands is expected to be maximal \cite{taillefer2009fermi}. However, the nonlinearity is observed over a wide doping range rather than being solely confined to the vicinity of the Lifshitz transition, which suggests that it does not originate from nonparabolicity alone. 

Even if non-parabolicity or interactions in the strange metal are the origins of the anomalous nonlinearity, it is interesting to note that the nonlinear response evolves smoothly through $T_c$. %In this regard, there is naturally another conducting state of matter that generically shows large THz range nonlinearities and that is superconductivity itself. 
Superconductivity itself can be a source of THz nonlinearities, as the driving $E$-field can break Cooper pairs. A distinctive feature of the data in Fig. \ref{Comparison}(g) is the correlation in the temperature and doping of the region where superconductivity is found with the region where the largest and most robust nonlinearity is observed. Note that this persists even in the normal state; i.e., LSCO films with higher $T_c$ also exhibit a larger nonlinearity, which extends to higher temperatures. One might understand this if there were local superconducting correlations that persist to temperatures much higher than the global $T_c$ and one is seeing the effects of their nonlinearities.

It is also noteworthy that the nonlinear response is smaller on the underdoped side as compared to their overdoped counterparts for similar $T_c$. This makes it unlikely that the physics of the pseudogap state is contributing significantly to the nonlinear response.  For example, the observation of an anomalous Nernst effect~\cite{wang2006nernst} in the normal state of the cuprates has been attributed to vortex excitations. While such an interpretation is not definitive~\cite{behnia2009nernst}, the smaller anomalous THz nonlinearity in the underdoped regime likely rules out correlations with the anomalous Nernst effect.

On balance of the various mechanisms, the totality of the data would appear to favor the explanation that the large nonlinearity arises from strong normal state inelastic scattering, but explicit calculations and more investigation needs to be done.  Even in the case that the nonlinearity is caused by essentially normal state physics, the correlation of the effect with superconductivity may be explained by the fact that the normal state correlations that are important for superconductivity may be the same as those that cause large THz nonlinearity.  For instance, Ref. \cite{cooper_anomalous_2009} has emphasized that the resistivity extrapolated to zero-field assuming a quadratic field dependence of the magnetoresistance implies that the $T$-linear regime grows wider and ``fans out" with decreasing temperature. This expansion of the $T$-linear region at low temperatures coincides with the superconducting dome at low temperatures and the region of superconductivity fluctuations above $T_c$.  Such behavior contrasts with what is expected in other nominally quantum critical systems \cite{custers2003break,paglione2003field}, where a V-shaped region of transport anomalies might be expected.  In a decomposition of the resistivity where different resistive contributions are added, the coefficient of the $T$-linear resistivity scales with $T_c$ going to zero near the edge of the superconducting dome on the overdoped side.  This may imply that the interaction causing anomalous energy-dependent scattering could also cause superconducting pairing.  Irrespective of the precise form of the resistivity, the tendency for resistivity to increase at rates faster than T$^2$ will give increased nonlinearity as discussed in Ref. \cite{kryhin2024strong}.

Qualitatively similar effects have also been observed in thin films of nearly optimally doped Bi$_2$Sr$_2$CaCu$_2$O$_{8+\delta}$ (Bi2212) ($T_c \approx$ 80K) (see \textit{Supplementary Information}), where we also observe large nonlinear response at low temperatures which smoothly evolve with increasing temperature without any discontinuity at $T_c$.  This indicates that the present observations are generic to the cuprates and not specifically related to LSCO. 

\subsection*{Experimental results on relaxation rates}

\begin{figure*}
    \centering
    \includegraphics[width = \textwidth]{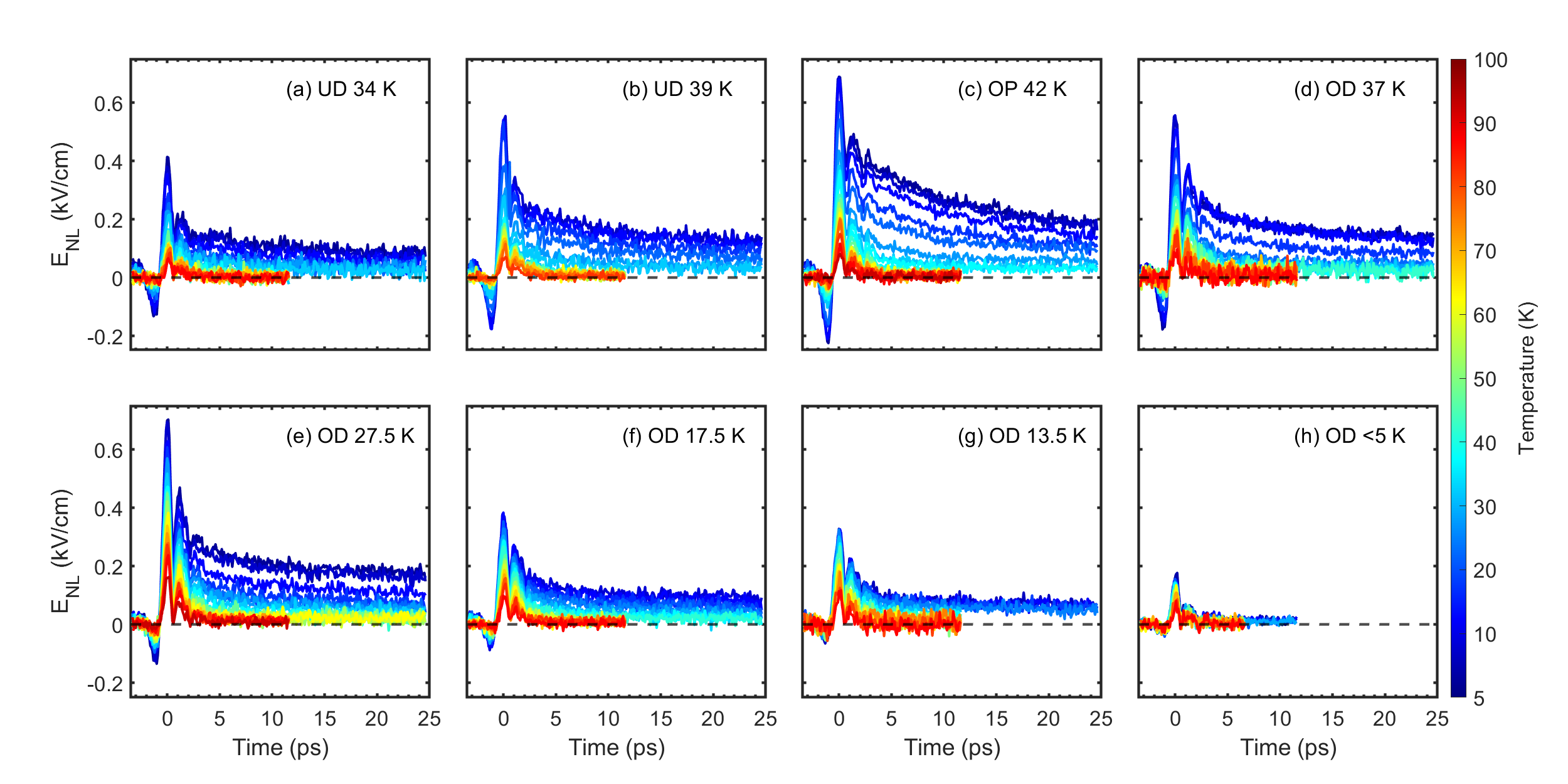}
    \caption{The temperature-dependent THz pump-probe response from 5 K up to 100 K.  In the low-temperature regime, we can parameterize the response into ``fast" and ``slow" decays, consistent with Rothwarf-Taylor dynamics in superconductors. This long-lived component persist to temperatures slightly above $T \gtrsim T_c$ as indicated by the offset from the $E = 0$ dashed line in the long time limit. In the high-temperature regime, only the ``fast" component is present in the pump-probe response with no offset observed.}
    \label{allPP}
\end{figure*} 

In Fig.~\ref{allPP}, we present the temperature-dependent THz pump-probe response for all samples measured. We use a strong-pump, strong-probe regime to allow for a direct comparison to the 2DCS response. The decay of the pump-probe response is a measure of the energy relaxation rate~\cite{barbalas2025energy,burt_unified_1990,yuen_differencefrequency_1982,rustagi_dispersion_1984}.  The signal originates in the same response that gives the strong horizontal streak in the 2DCS spectra in Fig. \ref{2D}.  We perform a ``1D" scan of $\tau$ (here at $t=3.3$ $ps$) as it is much faster than a 2D scan.  Although the 2D plots show strong $R$ and $NR$ peaks, they are broad in frequency and quickly decaying in time.  As shown in Fig. \ref{2D}, the long  $\tau$ \sout{scan} $PP$ signal is not influenced by the spectral features underlying the $R$ and $NR$ responses.  Accordingly, fitting a 1D scan yields the same result as fitting Lorentzians to the frequency scans along the $\omega_\tau$ direction if the 2D signal is properly rephased \cite{barbalas2025energy} and  the energy relaxation rate is weakly energy dependent, as it does at these temperatures. The early time response near $\tau \sim 0$ in Fig. \ref{allPP} is dominated by a coherent artifact within the overlap of the pump envelope that is likely due to parts of the sweep where $B$ comes before $A$, but for $\tau \gtrsim 4 $ ps, a fast exponential decay in the nonlinear response is observed.

We first briefly consider the superconducting state. The pump-probe decay is well-described by the phenomenological Rothwarf-Taylor (RT) model~\cite{rothwarf_measurement_1967}, which describes energy relaxation in BCS superconductors using a coupled quasiparticle-phonon model. After the quasiparticle and phonon baths thermalize individually, the rate of energy leaving the quasiparticles is limited by the emission of low-energy ($\hbar\omega<2\Delta$) phonons, termed the phonon-bottleneck. Evidence of this rate-limiting step is seen clearly in Fig.~\ref{allPP} for $T < T_c$, where the pump-probe response does not relax to 0 but rather a quasi-equilibrium situation is observed up to at least 25 ps after the initial pump pulse. After corrections due to internal reflections of the THz pump pulse within the sample are made, one can see this quasi-equilibrium state is unchanged up to at least 75 ps, suggesting a very long time constant. As the RT dynamics are governed by the phonon-phonon processes, the relative timescale for this phenomenon is typically  $\tau \sim 100-1000$ ps range~\cite{matsunaga_nonequilibrium_2012}. The observation of RT dynamics in the superconducting state has been observed across the cuprate family using optical range pump-probe experiments, including in LSCO~\cite{kusar_systematic_2005}, YBCO~\cite{averitt_nonequilibrium_2001,demsar_superconducting_1999, segre_photoinduced_2002,gedik_single-quasiparticle_2004,chwalek_subpicosecond_1991,hinton_rate_2016}, BSCCO~\cite{kaindl_dynamics_2005,toda_quasiparticle_2011}.

In the normal state, this pump-probe  response is governed by the rate that energy leaves the electronic ensemble~\cite{barbalas2025energy}.  In conventional systems at low temperatures, this is given by the coupling to the acoustic phonons~\cite{allen1987theory,glorioso2022joule}.  To extract the energy relaxation rates, we fit the pump-probe response to a phenomenological model of an exponential decay plus a constant,
\begin{equation}
    E_{NL}(\tau) = A \,\textrm{exp}[-2\pi\Gamma_E \, \tau] + C,
\end{equation}
where $\Gamma_E$ corresponds to the fast decay due to energy relaxation and $C$ corresponds to the contribution from a long-lived quasi-equilibrium state. Since the lifetime of the quasi-equilibrium state in the superconducting phase is much longer than the experimental time window, treating the quasi-equilibrium state as a constant is reasonable.   The constant is effectively zero in the normal state and the response given by simple exponential decay.

\begin{figure*}
    \centering
    \includegraphics[width = \textwidth]{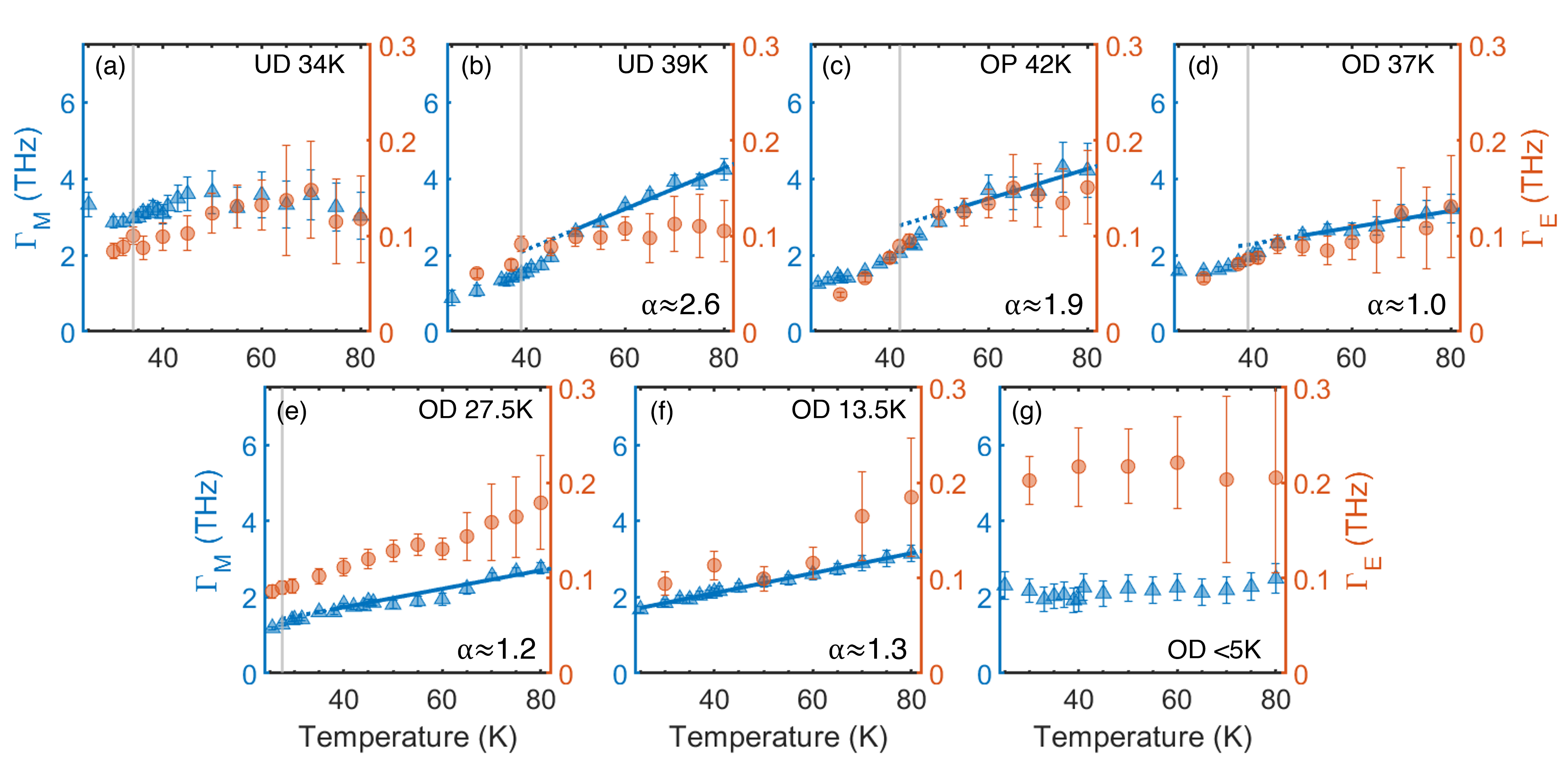}
    \caption{Comparison of $\Gamma_M(T)$ from time-domain THz spectroscopy (left, blue) with the electronic $\Gamma_E(T)$ (right, orange) from the THz pump-probe experiments. The black dashed line indicates $T_c$ for each sample. $\Gamma_M(T)$ remain much larger than the $\Gamma_E(T)$ for temperatures above $T_c$. Below $T_c$, extraction of the $\Gamma_M(T)$ via Drude fitting becomes unreliable and only serves as a qualitative reference. For all but the most UD and OD samples, the scattering rate is consistent with the Planckian form $\Gamma_0 + \alpha kT/h$ with $\alpha =$ 2.6, 1.9, 1.0, 1.2, and 1.3 for the data in panels (b)- (f).}
    \label{Rate_comparison}
\end{figure*}

In Fig.~\ref{Rate_comparison} we show the extracted energy relaxation rates $\Gamma_E$ plotted alongside the Drude $\Gamma_M$ as a function of temperature.  $\Gamma_M$ comes from fits to the linear response time-domain THz spectroscopy (TDTS) spectrum as measured routinely by our group in the THz range~\cite{bilbro2011temporal}. (See \textit{Supplementary Information} for sample fits).  For all samples except for the most underdoped, the complex conductivity can be well-described by the Drude form (and even in the superconducting phase for very overdoped samples~\cite{mahmood_locating_2019}).  $\Gamma_M$ inferred from the width of the Drude response shows a temperature dependence roughly consistent with the ``Planckian" form i.e. $\Gamma_M(T) = \Gamma_0 + \alpha k T/h$ with $\alpha \approx 2$ for optimally doped samples.  Note that the width of the Drude peak from an optical conductivity experiment provides a direct measure of the current relaxation rate with none of the assumptions about effective mass or carrier density that is needed to interpret dc transport data.  That $\alpha$ can be measured in a model independent fashion from optics and found to be approximately twice the values arrived at from dc transport~\cite{legros_universal_2019} is an old result from optics~\cite{quijada1999anisotropy,liu2006drude}, but one that is not widely appreciated.   Recent newer work confirms this dependence~\cite{van2022strange}.

We focus on the energy relaxation at temperatures above the superconducting fluctuation region~\cite{bilbro2011temporal} and in the temperature regime that the momentum relaxation has its linear dependence on temperature. While all samples show the momentum and energy relaxation rates increasing similarly with $T$, $\Gamma_M$ is significantly larger than the $\Gamma_E$ for all samples and temperatures.   This indicates that the scatterings that cause resistance are largely not the same as those that carry energy away from the electrons.   A large temperature dependent $\Gamma_M$ can arise from momentum non-conserving electron scattering, which removes energy from the electron ensemble. The long time scale after laser pumping has been observed previously in 1.5 eV pump-photoemission experiments~\cite{yang_inequivalence_2015}, but, as discussed above, assumptions about rapid thermalization and accounting for phonon-bottleneck effects have to be made. We would also note that the rates extracted from optical-pump optical probe data on similar LSCO samples are faster than the ones we extract here ~\cite{kusar_systematic_2005}.  To ensure that the measured energy relaxation reflects intrinsic scattering processes, the THz pump-probe response was characterized as a function of pump fluence (see the \textit{Supplementary Information} for more details). We observe no fluence-dependence of the energy relaxation rates in the normal state, indicative that our measure of the energy relaxation reflects an intrinsic quantity.

\subsection*{Discussion on relaxation rates}

Our data shows that there are interactions beyond electron-phonon which dominate the physics of the cuprates.  It is important to note the temperature dependence of $\Gamma_E$.  In the conventional theory~\cite{allen1987theory}, the primary channel for energy relaxation at low temperatures is coupling of quasiparticles to acoustic phonons. As mentioned above, it is generally expected that $\Gamma_M > \Gamma_E$, as momentum loss can also occur through electron-electron umklapp processes and disorder scattering. At low temperatures, $\Gamma_E(T)$ is predicted to follow the form $\Gamma_E = 12 \zeta(3) \lambda \Big[\frac{k_B T_e}{\hbar \omega_D}\Big]^3 \omega_D$, where $\zeta$ denotes the Riemann zeta function, and $\omega_D$ is the Debye frequency~\cite{allen1987theory}. $\lambda$ is the dimensionless electron-phonon coupling constant.  In the phonon equipartition regime at higher temperatures, electron-phonon scattering crosses over to being predominantly elastic, causing $\Gamma_E(T)$ to decrease as $1/T$ (reflecting the temperature dependence of the fermionic heat capacity). This leads to a peak at a temperature that is a fraction of the Bloch-Gr{\"u}neisen temperature. Notably, this temperature scale coincides with the point at which the phonon contribution to resistivity in a conventional metal is expected to transition to a $T$-linear dependence. Similar considerations hold for optical phonons at temperatures above their characteristic energy scale~\cite{glorioso2022joule}.   The fact that our energy relaxation rate is an increasing function of $T$ indicates that the bath that the electrons are loosing energy to is not in its equipartition regime.

We believe these observations together (the overall scale of $\Gamma_E$ as compared to $\Gamma_M$ and its temperature dependence) rules out scattering off phonons as a source of $T$ linear resistivity.  Although it was traditionally regarded that phonons did not play a major role in the physics of the cuprates, it is in fact the case that at least not too high of temperature (but for $T \gg T_{BG}$ the resistivity of cuprates can be fit with a linear in $T$ scattering rate of the form $h \Gamma_M = 2 \pi \lambda k T $~\cite{allen2001kinky,hwang_linear2019,das2024role} from the well-established Bloch-Gr{\"u}neisen theory.  With not even too high values for $\lambda$ a ``Planckian" scaling can be reached.   However this would be the temperature regime where $\Gamma_E$ is a decreasing function of temperature, which is not observed in our data.

One question is then;  what other scenarios does our data rule out?   Does it rule out $T$-linear resistivity arising from coupling to collective bosonic modes that are essentially electronic in character? Can such modes, in principle, remove energy from the electronic system despite the fact that such modes are themselves made from electrons?   If so, then again the fact that  we find $\Gamma_E \ll \Gamma_M$ rules out such modes as being a source of linear in $T$ momentum scattering.  Guo et al. considered a slightly different question \cite{guo2025phonon}, and studied how electrons in a quantum critical metal with a nematic like collective mode loses energy to acoustic phonons, treating the phonons as a bath and characterizing the rate of energy relaxation via various symmetry‐allowed couplings.  They found a rich sequence of temperature‐dependent powerlaws in $\Gamma_E$, which arise from the different possible coupling channels (electron-phonon, boson-phonon) that may be compared to data.   Future refinements to this technique may allow these comparisons.

Going forward, it will be important to characterize the temperature dependence of the power laws in $\Gamma_E$ more precisely to compare to recent predictions~\cite{guo2025phonon}.   We hope that even shorter THz pulses, possibly using air plasmas, will allow for the measurements of faster relaxation times and perhaps the relaxation of anisotropic B$_{1g}$ Fermi surface deformations.  It will also be interesting to do experiments with narrow band pulses that may enable a more detailed analysis of conversion efficiencies of the nonlinear processes as done in graphene~\cite{hafez2018extremely} and support a quantitative comparison with theory to precisely determine the underlying mechanism of the nonlinearity.

\section*{Methods} \label{Methods}

\subsection*{THz 2D Coherent Spectroscopy}

The information provided by nonlinear THz spectroscopy is particularly useful as one can probe properties of materials that may be otherwise invisible to linear response conductivity. To study the nonlinear response in a 2DCS/pump-probe experiment, intense THz pulses were generated in a pair of LiNbO$_3$ crystals using an amplified 800 nm Titanium:sapphire laser with $\sim 3$ mJ incident on each crystal in the tilted-pulse-front scheme~\cite{hirori2011}. Using a two-pulse measurement scheme ($E_A$ and $E_B$) with a differential chopping arrangement the response is measured from $E_A$ and $E_B$ individually and when both are incident on the sample $E_{AB}$. The resulting nonlinear response due to both pulses is then obtained via $E_{NL} = E_{AB} - E_A - E_B$. The transmitted THz pulses were detected with standard electro-optic methods using ZnTe (for the single-pulse transmission experiments) or GaP (for all other two-pulse measurements) detection crystals. More details on the setup can be found  elsewhere~\cite{mahmood_observation_2021}.

The pump-probe signal was measured by keeping $E_A$ fixed at a point in time ($t$) and varying the delay $\tau$ between $E_A$ and $E_B$. The resulting nonlinear response is then the changes in the response due to $E_B$, correcting for nonlinear effects already present due to $E_A$ only. In this sense, the nonlinearity reported is uniquely due to the presence of both pulses $E_A$ and $E_B$, eliminating any background signal present in either channel individually. 

\subsection*{Sample growth and characterization}

The La$_{2-x}$Sr$_x$CuO$_4$ thin films used for the study had dopings ranging from the underdoped region ($T_c = 7$ K) into the non-superconducting overdoped region ($T_c \approx 0$ K). The c-axis oriented single crystal films were grown by molecular beam epitaxy (MBE) on LaSrAlO$_4$ (LSAO) substrates to a thickness of 24 nm. The films were grown via MBE and characterized using RHEED and XRD verifying high crystalline order. After annealing, the films were also characterized using 4-probe resistivity measurements and mutual inductance to measure $T_c$. We measure samples ranging in doping from mildly underdoped to extremely overdoped where superconductivity is suppressed. Using the empirical formula $T_c/T_{c,max} \approx 1-82.6(p-0.16)^2$, the nominal doping for the films can be extracted~\cite{presland1991general}.

\subsection*{Momentum relaxation rates}

To extract the momentum scattering rate from the THz complex conductivity, data was fit to the Drude model. The raw conductivity curves are shown in \textit{Supplementary Information} for reference. In all of the superconducting samples we see the expected $1/\omega$ at the lowest temperature.  At high temperatures we observe the expected conductivity for a disordered metal with a large scattering rate.  We fit the data with a Drude term and for $T \leq T_c$, a contribution from superconductivity $iS_{\delta}/\omega$. The change in the real part of the conductivity from the superconducting to the normal state is small for the overdoped samples  \cite{mahmood_locating_2019}.  For quantitative comparison, we focus on the normal state, between ($T_c+10$)K and 80K, which is above the superconducting fluctuation regime \cite{bilbro2011temporal}. The data demonstrated good agreement for all samples with the Drude fits.   Representative fits in the normal state between ($T_c+10$)K to ($T_c+50$)K are shown in \textit{Supplementary Information}. 

\section*{Acknowledgements}

We would like to thank Yufan Li for depositing the gold film and S. Kryhin, S. Sachdev, D. Tanner, D. van der Marel, and P. Volkov for helpful conversations.

\section*{Declarations}

\textbf{Funding}: The project at JHU was supported by the NSF-DMR 2226666 and the Gordon and Betty Moore Foundation’s EPiQS Initiative through Grant No. GBMF9454.  NPA had additional support from the Quantum Materials program at the Canadian Institute for Advanced Research.  Work at Brookhaven National Laboratory was supported by the DOE, Basic Energy Sciences, Materials Sciences and Engineering Division. X. H. and I.B. were also supported by the Gordon and Betty Moore Foundation’s EPiQS Initiative through grant GBMF9074.

\textbf{Conflict of interest/Competing interests}: The authors declare no competing interest or conflict of interests.

\textbf{Data availability}:  All the data used in the plots displayed in this paper and the \textit{Supplementary Information} are available from the corresponding authors upon reasonable request. 

\textbf{Code availability}:  The analysis codes used to process the data are available from the corresponding authors upon reasonable request.

\textbf{Author contribution}: DB and DC conducted the THz nonlinear pump-probe experiments and performed data analysis. JL, RR, FH and AL performed the characterization of the optical conductivity and assisted with the nonlinear THz experiment. XH and IB grew and characterized the LSCO thin films. HR grew the Bi2212 film.  DC, DB, and NPA wrote the manuscript and all authors edited it.  NPA directed the project.

\bibliographystyle{apsrev4-2}   
\bibliography{ref}

\clearpage
\includepdf[pages=1]{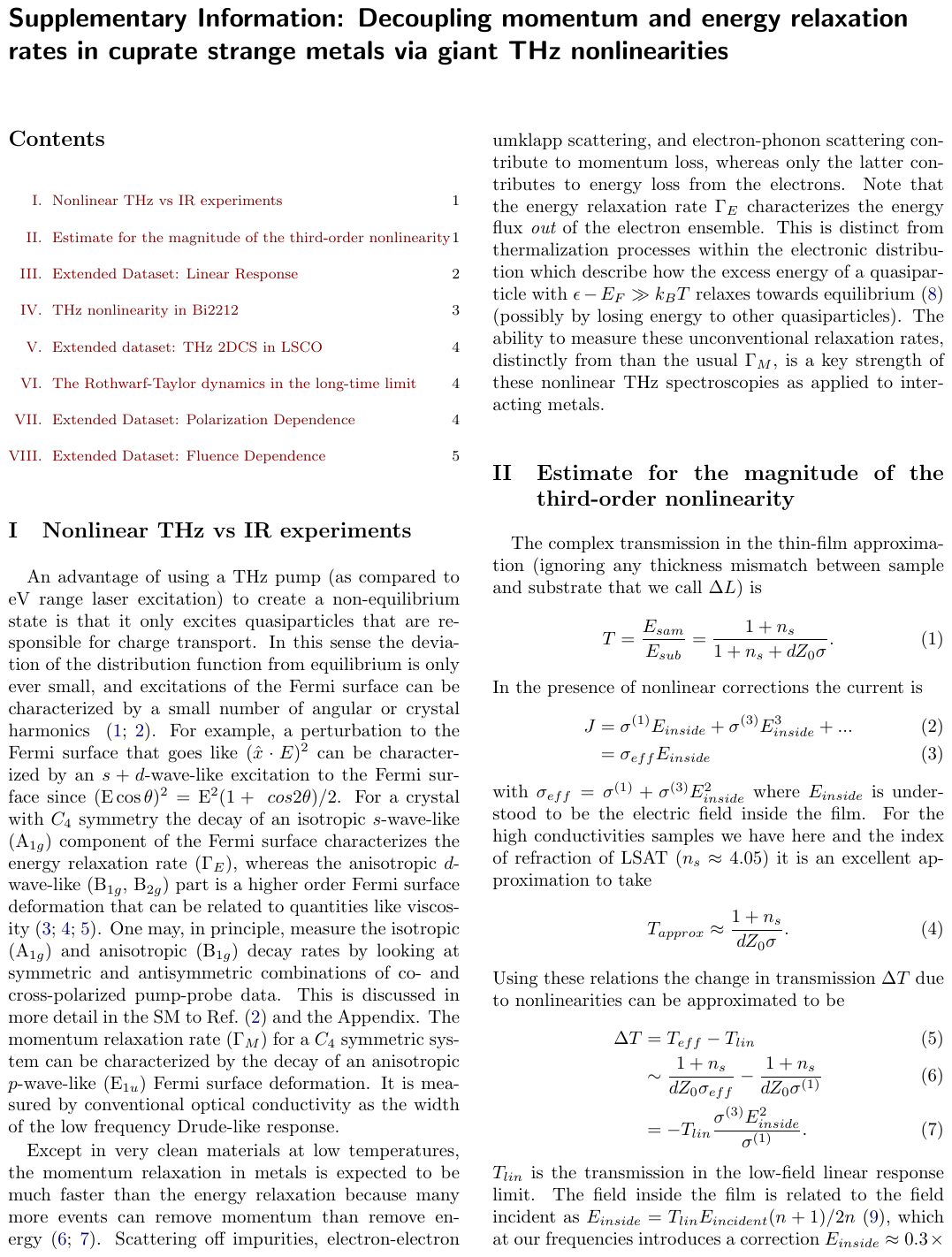}
\clearpage 
\includepdf[pages=2]{SI.pdf} 
\clearpage 
\includepdf[pages=3]{SI.pdf} 
\clearpage 
\includepdf[pages=4]{SI.pdf} 
\clearpage 
\includepdf[pages=5]{SI.pdf} 
\clearpage 
\includepdf[pages=6]{SI.pdf} 

\end{document}